\documentclass[a4paper,11pt]{article}

\usepackage[utf8]{inputenc}
\usepackage{natbib}
\usepackage[T1]{fontenc}
\usepackage[french, english]{babel}
\usepackage[hyphens]{url}
\usepackage{csquotes}
\usepackage{fourier}
\usepackage[top=3cm,left=3.5cm,right=3.5cm,bottom=4cm]{geometry}
\usepackage{array}
\usepackage{multirow}
\usepackage{float}
\usepackage{hyperref}
\usepackage{graphicx}
\usepackage{array}

\usepackage{sectsty}
\subsectionfont{\normalfont\itshape}

\usepackage{endnotes}
\let\footnote\endnote

\makeatletter
\def\enoteformat{%
  \rightskip\z@ \leftskip\z@ \parindent=1.8em
  \leavevmode{\setbox\z@=\lastbox}\llap{\theenmark.\enskip}}

\usepackage{color}

\newcounter{example}

\usepackage{fancyvrb}

\usepackage{amsmath}

\usepackage{bm}

\usepackage{authblk}

\newcommand{\absdiv}[1]{%
  \par\addvspace{.5\baselineskip}
  \noindent\textbf{#1}\quad\ignorespaces}

\title{Mining User Queries with Information Extraction Methods and Linked Data}
 
\author{Anne Chardonnens\thanks{Corresponding author}, Ettore Rizza, Mathias Coeckelbergs, Seth van Hooland \\ 
Université libre de Bruxelles (ULB)\\
ReSIC Research Center\\ 
Information and Communication Science Department\\ 
Avenue F.D.\,Roosevelt\,50 -- CP\,123 --
B-1050 Brussels, Belgium\\
\texttt{\{anchardo,erizza,mcoeckel,svhoolan\}@ulb.ac.be}
}

\makeatother

\begin{document}
\maketitle

\begin{abstract}
\absdiv{Purpose} Advanced usage of Web Analytics tools allows to capture the content of user queries. Despite their relevant nature, the manual analysis of large volumes of user queries is problematic. This paper demonstrates the potential of using information extraction techniques and Linked Data to gather a better understanding of the nature of user queries in an automated manner.

\absdiv{Design/methodology/approach} The paper presents a large-scale case-study conducted at the Royal Library of Belgium consisting of a data set of 83 854 queries resulting from 29 812 visits over a 12 month period of the historical newspapers platform BelgicaPress. By making use of information extraction methods, knowledge bases and various authority files, this paper presents the possibilities and limits to identify what percentage of end users are looking for person and place names. 
\absdiv{Findings} Based on a quantitative assessment, our method can successfully identify the majority of person and place names from user queries. Due to the specific character of user queries and the nature of the knowledge bases used, a limited amount of queries remained too ambiguous to be treated in an automated manner.   
\absdiv{Originality/value} This paper demonstrates in an empirical manner both the possibilities and limits of gaining more insights from user queries extracted from a Web Analytics tool and analysed with the help of information extraction tools and knowledge bases. Methods and tools used are generalisable and can be reused by other collection holders. 
\absdiv{Keywords} User query; Query Classification; Digital Libraries;
Cultural Heritage 
\absdiv{Paper type} Case study 
\end{abstract}

\section{Introduction}

Both policy makers and the public are increasingly regarding libraries, archives and museums as content and service providers who operate in the same market as commercial information providers. This situation is reflected in the adoption of the common definition of the \emph{quality} of information systems and services by ISO within the cultural heritage sector, which focuses
on the \enquote{fitness for purpose} \citep{iso2005, Boydens1999}. This interpretation of quality refers to the idea of self-regulating markets where demand directly influences supply as consumers are empowered to decide what information is of use \citep{suominen2007problem}.

Within this context, cultural heritage institutions have been making use of Web Analytics tools to quantify the interaction between their collections and end users. The dashboards of popular tools such as Google Analytics do provide useful features to understand how many end users interact with a website, where they come from, how long
they stay or with which specific web pages they interact. However, this approach does not provide a detailed analysis of how patrons interact for example with information retrieval systems. Other tools and methods are required to fill this current gap.

This paper seeks to aid in developing such methods and aims and, more in particular, to demonstrate how automated methods can help a cultural heritage institution to interpret a large corpus of user queries. For this purpose, the paper presents a case study from the Royal Library of Belgium.
Launched in 2015, Belgica Press\footnote{\url{http://opac.kbr.be/belgicapress.php}} provides online access to more than two million pages of digitised Belgian newspapers spanning the period 1831-1950\footnote{Parts of the corpus, subject to copyright laws, are exclusively available within the library}. The user interface offers functionalities such as full text searching across the OCR'ed pages. Other search parameters include time ranges, specific dates, newspapers titles and languages (French, Dutch or German).

Apart from which periods and what specific journals are consulted,
the library also wishes to know to what extent end users perform queries
based on a personal name or a place name. These two types of Named
Entities (NE) are presumably considered the most frequent in historical
corpora of French and Dutch newspapers \citep{neudecker2016open}. Indeed,
the last few years have seen an increasing interest in Named Entity
Recognition (NER) and Linked Data to semantically enrich metadata (see for instance the experiments carried out by the National Library of the Netherlands \citep{van2015semantic}. Beyond the hype of such practices, which often stimulated the development of new tools and
methods as an end itself, one might wonder whether the efforts made
by institutions to develop these new ways to access documents actually
meet users' needs. Usage data appears as an opportunity to assess
whether the offer meets the demand.

Concretely, this paper focuses on the recognition of personal names
and places contained within user queries by making use of information
extraction methods, knowledge bases (KBs) and various authority files. In
doing so, this paper will try to answer the following question: \\
 \emph{How can Linked Data help to identify the presence of personal
names and place names mentioned in user queries?}

\noindent The paper starts with a literature overview of relevant
research on the aggregation and interpretation of usage data in the
cultural heritage sector and the application of Information Extraction
Methods, after which the case study of the Royal Library in Belgium
and the related methodology is presented in detail. An overview of
the manual annotation and the data extraction process is then presented,
followed by results and discussion. The conclusions focus on how the
methods and tools are generalisable for other collection holders and future work.

\section{Related Work}
Online user behaviour has been increasingly analysed in the cultural
heritage field, especially since the launch of Google Analytics in
2005. As highlighted by \citet{kelly2014assessment}, Web Analytics can lead to developing enhancements to the architecture, metadata or content of a digital
library to improve the user experience. Various methodologies
and metrics have been developed to fit archive and library website
specificities. For example, \citet{fagan2014suitability} illustrated how commercial key performance indicators can be adapted to an academic library environment. Still, beyond assessment tools helping to collect user experience,
\enquote{additional tools for automating and analyzing this data
are still needed to make it a widespread practice {[}for archives{]}}
\citep{kelly2017altmetrics}.

Thus, as underlined by \citet{zavalina2014understanding}, very few studies
examine the content of the user search queries and other relevant
log files. \citet{ceccarelli2011improving} used Europeana query logs, but more as means for developing assistance functionalities such as a query recommender system than as objects of study per se. Likewise, \citet{dijkshoorn2014using} considered the log files from the Rijksmuseum as an aid for combining user queries with external vocabularies published as Linked Data, in the attempt
to diversify search results. In both cases, no text mining methods
were used, requiring substantial manual work and interpretation.
In contrast, \citet{zavalina2007collection} mentioned, in the context of the IMLS Digital Collection query logs, that some processing of the queries (truncating plural forms, excluding stopwords such as prepositions, etc.) has been done before the categorisation and semantic matching with a controlled vocabulary. However, the whole process, including the extraction of all query strings from the log files, was done manually on a corpus containing fewer than 1 000 queries.
This example highlights the potential of computational methods to save time and be applicable to larger datasets using a script to semi-automatically process the data and automatically extract information. 

Within the more specific area of online digitised newspapers, \citet{dhq} presented a case study from the \emph{Historische Kranten} project. Before focusing on the potential of NER and linked
data to enrich multilingual archives metadata, they evaluated user demands. For this purpose, they tracked individual queries over a 4-year period. Their findings revealed that, according to the ten most popular keywords, locations are especially favoured. Although promising, the analysis is not further developed. \citet{gooding2016exploring} performed an overall analysis of the information behaviour of users of Welsh Newspapers Online Website. Using in a complementary way the possibilities offered by Google Analytics and web server logs, he observes that the first one is not tailored
for academic research and provides a weaker source for in-depth analysis,
due to the opacity of data processing and the impossibility for the
user to export raw data. The web logs allow him to identify the most viewed newspaper titles, the most viewed decades and the most commonly viewed page numbers. At last, he observed that “over half of page views are dedicated
to interacting with the web interface rather than the historical sources”.
Although this work fills a gap in the literature, the paper offers
room for experimentation in the analysis of the content
of user queries themselves.

Progress can be made by building on experience gained in the broader
field of web query classification. Query labelling is known to be
arduous due to the nature of web queries, which are usually short
in length, grammatically unstructured, and semantically ambiguous
\citep{Alasiry2015a}. The development of automatic classification
techniques has been stimulated by the 2005 KDD Cup (held during the
ACM Conference on Knowledge and Data Discovery): a competition to
automatically classify 800 000 search queries, without training data. While some authors, like \citet{cao2009context}, made use of contextual
information (both previous queries within the same session and retrieved
results) to classify queries, \citet{Beitzel2007}, being limited
by operational restrictions, showed the possibility to \enquote{topically
classify a significant portion of the query stream without requiring
external sources of information}. To do so, they used a method combining
manual classification, supervised learning classification, and rule
classification. Work has also been done using external information
sources such as the Wikipedia structure. For instance, \citet{khoury2011query} exploited the structure to build a general-domain query classification
system, by matching the query words to Wikipedia titles.

Finally, web query classification has been associated with Named Entity Recognition and Classification (NERC) techniques \citep{pacsca2007weakly, guo2009named}.
The task consists of assigning an entity type (e.g. person, location,
or company) to all entities identified within the search queries.
In the context of modern web search engines, NE stored in Knowledge Graphs are used as result pages for entity-centric search queries \citep{tonon2016contextualized}. Such systems underline the need to develop
entity disambiguation and entity types ranking \citep{demartini2010finding,tonon2013trank,van2014linked,shen2015entity}, while highlighting the potential for mining user queries from our corpus.

\section{Methodology}
As aforementioned, the Royal Library of Belgium is interested
in understanding the information needs of its patrons regarding the
historical newspapers published as \emph{BelgicaPress}. Its aim is
to be able to quantify the presence of personal names and place names
in user queries.

From both a conceptual and an empirical perspective, it is difficult to define formal and exclusive categories such as person or place names for queries. For certain queries, the line will certainly be blurry. From a Natural Language Processing (NLP) point of view, the distinction for example between a family name, a place name or the name of an organisation can be very problematic. In the Brussels
context, the string of characters \enquote{Wiels} can refer either to a family name, to a geographical reference of the location where a famous beer brewery led by that family was installed, the brewing company or the current art centre housed in the former brewery. 



In order to implement this research question, we first extracted a one-year corpus of user queries, from \enquote{raw} data collected by Piwik, the open-source Web Analytics tools used at the library. Secondly, we manually annotated a sample of 1 000 user queries to create our
Gold Standard Corpus (GSC) and obtain a first idea of the amount of NE contained in the queries based on a manual analysis. The relatively poor results obtained through 7 NER services led us to develop our own script to extract personal names and (Belgian) place names. Using Natural Language Processing (NLP), we reconciled potential NE from the queries against several KBs and authority files published as Linked Data. This process has resulted in a satisfying F-score (evaluation section), semantic enrichment of the queries and enlightening results
(results section). All of the above mentioned steps of the methodology are discussed in detail in the following sections. 

\section{Data}
User queries of Belgicapress are aggregated by two different sources: the log files stored by the database management system and the \enquote{raw} data collected by Piwik in the context of the MADDLAIN project. 

Considering the possibilities and limits offered by these two sources of data, we have chosen to use Piwik to create our dataset. Beyond strictly logistical reasons (the institution had stopped collecting the log files during the year 2016), our choice was guided by the fact that the Piwik data do not require the pre-processing steps required
by the log files to recreate the sessions of each visitor. Moreover, by combining IP addresses and HTTP cookies to identify visitors, the
Piwik data provides more accurate visitor numbers than log files. By way of illustration, for an equal test period (October 2015), almost 18\% of distinct visitors identified by Piwik were not recognised as such in the log files (Log files: 1071 visitors identified via IP addresses, Piwik data: 1298 visitors identified via cookies and IP addresses).

The dataset covers 12 months (from January 1 2016 to January 1 2017) and contains five elements that have been extracted from the Piwik database: 
\begin{itemize}
\item the visitor ID; 
\item the visit ID; 
\item the timestamp: day, hour, minute and second of the action; 
\item the URL, whose user queries can be parsed ; 
\item the custom variables containing additional information about the browsing
behaviour: full-screen mode activation, full view of a newspaper, a click in the list of publications, etc. 
\end{itemize}
These five elements can be found back in an example where \enquote{Bruxelles} is the user query (see Table 1).

\begin{table}
\begin{center}
\begin{tabular}{l p{10cm}} 
\hline
 Visitor ID & B9A3E4383488480  \\
\hline
Visit ID & 650082  \\
\hline
Timestamp & 2016-01-01 08:12:45  \\
\hline
URL & \url{http://opac.kbr.be/pressshow.php?adv=1&all_q=BRUXELLES&any_q=&exact_q=&none_q=&from_d=&to_d=&per_lang=&per=&lang=FR 
} \\
\hline
Custom variables &  pageview\_select\_page 14-09-1894, Ed.1 p.4 \\
\hline
\end{tabular}\caption{Example Piwik Data (the query part has been capitalised).}  
\end{center}
\end{table}

The Piwik data require computational methods to be exploited: given that the user query needed for analysis is contained in an URL and not presented in a structured way\footnote{It has to be noted that Piwik offers functionalities to track internal search keywords and obtain them more easily. These functionalities have not been implemented in the context of our project, but could potentially facilitate the process explained in this paper.}, URLs have first to be parsed. More specifically, this means that the relevant data are identified within the URL and then automatically extracted in a structured file (the method to pre-process the data is described in a similar case study, see \citet{chardonnens2017text}).

At the end of the pre-process, the dataset contains a total number of 83 854 queries, among which 52 547 distinct queries. A little less than 30 000 visits on BelgicaPress website (29 812) are at the origin of these 83 854 queries: this leads us to an average of 2.87 distinct queries per visit (standard deviation: 4.1 and median: 1).
The minimum is 1 and the maximum is 107 queries per visit. 
The number of tokens per query ranges from 1 to 64 (average: 1.8,
standard deviation: 1.1, median: 2). Moreover, it has to be noted that
98 percent of the dataset (82 279) contains 5 tokens at most.

\section{\label{sec:section5}Creation of a Gold Standard Corpus}

To have a first look at the data, a manual analysis was required.
The aim was to examine which proportion of queries actually contains
personal names (PER) and place names (LOC), by annotating a representative
sample of 1 000 randomly selected queries. In addition, the result
of this annotation will produce a GSC, which will
subsequently be used to evaluate the outcome of the automated extractions.

These two basic categories - PER and LOC - have been extended to face ambiguity issues during the annotation task, resulting in five temporary categories: 
\begin{itemize}
\item LOC for locations in a broad sense: any geographical location corresponding
to a place, be it a municipality, a country name or even a subway
station (e.g. \enquote{Horta station}). 
\item PER for a full name, a last name or just a first name (e.g. \enquote{Leopold
II}). When relevant, the presence of a full name (first name and
last name) was reported in an additional column. 
\item PER LOC for ambiguous cases where the entity may designate both a
place or a person (e.g. \enquote{général Jacques}, which turns
out to be at the same time the name of a Belgian soldier and a street
in Brussels). 
\item PER AMBIG for entities which vaguely look like a person's name, but
no known place or person can be associated with (e.g. \enquote{Tombek}). 
\item AMBIG for very ambiguous tokens: it could be an entity named PER or
LOC as well as another type or a common name (e.g. \enquote{stampe},
\enquote{valk}). 
\end{itemize}
Moreover, a general rule has been set: no overlap is allowed. Thus, \enquote{August
van Turnhout}, which is clearly a full name, will be annotated as
such, while \enquote{Turnhout} will not be annotated separately
as LOC, although it is a Belgian locality.

The annotation task intended to dive into the user's mind to try to
understand what he or she was looking for. This means that the trained annotators
performed the annotation using contextual data (the other search terms
entered during the same visit), to obtain information as accurate
as possible and reduce ambiguity. Thus, considered alone, the query
\enquote{Corbiere}, is quite vague. A glance at the previous and subsequent
queries (\enquote{de la Corbiere} and \enquote{Lacorbiere}) lets us assume that the
query probably refers to a PER entity (the French painter \enquote{Roger de
la Corbière}) and not \enquote{Corbières}, the Swiss location or the French wine of the same name. 

Once emptied of duplicates and inoperable texts (queries composed
only by numbers), the sample consists of 995 queries. 
These were manually annotated by two of the authors, with the help
of online search engines and databases such as Wikidata or Geonames.
At the end of the process, divisions of opinion were discussed with
a view to reaching a consensus. When no consensus was reached, each
author retained his initial annotation.

\begin{table}
\centering{}%
\begin{tabular}{|l|r|r|r|}
\cline{2-4} 
\multicolumn{1}{l|}{} & Annot. 1  & Annot. 2  & Consensus \tabularnewline
\hline 
PER  & 485  & 477  & 473 \tabularnewline
\hline 
PER\_AMBIG  & 18  & 25  & 17 \tabularnewline
\hline 
PER\_LOC  & 18  & 19  & 16 \tabularnewline
\hline 
LOC  & 317  & 317  & 313 \tabularnewline
\hline 
AMBIG  & 10  & 11  & 10 \tabularnewline
\hline 
\end{tabular}\caption{\label{tab:Results-of-the}Results of the double annotation}
\end{table}

At the end of the annotation task, a total of 849 entities were identified (see Table 2). Out of these entities, 829
(97,6\%) were classified in the same category by the two annotators. In spite of some dissonances, the consensus method resulted in a high degree of inter-annotator agreement, with a Cohen's kappa \citep{cohen1960coefficient} higher than 0.96 on a maximum of 1.

Of the 313 LOC resulting from the consensus, 225 (78\%) are located in Belgium, far more than the Democratic Republic of the Congo (9\%),
France (6\%), the Netherlands (5\%) and about fifteen other countries,
each of which has fewer than ten occurrences.
In addition, it is interesting at this stage to note that the 225
Belgian geographic references are mainly composed of names of municipalities
or municipal districts (86\%). The remaining 24\% are \enquote{Points
of Interest}, such as a particular building (\enquote{église
Saint-Paul}), names of subregions and provinces, forests
or rivers.

\section{Extracting Place Names and Personal Names}

This section describes the method developed to automatically extract
and categorise NE corresponding either to the type PER (person) or LOC (location). This method is designed to be generalisable, in the sense that cultural heritage institutions with limited human and financial
resources are able to reuse the script in similar contexts, namely in the extraction of places and personal names in short and unstructured texts, regardless of the language
of the corpus. In theory, our method is applicable to most Western languages, although we have only tested it on queries in French, Dutch and English. Hence, language-specific problems which do not occur in this family of languages, such as word separation problems as known from Thai and Chinese for example, might cause the need for specific attention, which goes beyond the scope of this paper. 

Like most information retrieval systems, NERC methods
can be broadly divided into three categories: rule-based
systems, machine-learning systems, and mixed methods \citep{nadeau2007survey}.
But if they can obtain results similar to those of a human being on
texts in structured language, such as press articles in English, these
methods prove to be faulty on short texts, due to their informal and ambiguous
syntax. These include NER for short text messages \citep{ek2011named},
tweets \citep{derczynski2015analysis, strauss2016results} and, of course, user queries in a search
engine \citep{cornolti2016piggyback}.

\noindent Preliminary tests performed on our GSC with seven NER/entity
linking web services (Rosette,
Dandelion,
Babelfy, 
TagMe, 
Dexter,
DBpedia Spotlight
and the Stanford NE Recogniser trained on English) 
led to poor results. The service which produced the most convincing
result, Rosette, correctly identified only 71 names of Belgian municipalities and municipal districts out of a total of 198 in the GSC and 128 full names of persons out of 141.

The web services are not adequate to deal with the singular
nature of the corpus, i.e., ambiguous, unstructured and very short
texts, which are less than three tokens per query on average. To fill this
gap, the decision was taken to internally develop a tool more adapted
to the specific nature of the queries. The tool consists of a Python script based on a minimum set
of gazetteers and hand-written linguistic rules \footnote{Although in development, the script is already accessible on GitHub
: \url{https://github.com/ulbstic/BelgicaPress}. }. The next two subsections briefly describe our pipeline: first, the
extraction of place names, then, the extraction of personal names;
both based on slightly different logic. 

\subsection{Place Names}

In this subsection, we describe the pipeline to automatically extract
location names. In order to limit noise in the results, we have decided
to work on a Belgian scale: the manual annotation described in section
\ref{sec:section5} has shown a significant presence of Belgian municipalities
and municipal districts among locations mentioned in the queries ---
which makes sense in a database of Belgian newspapers. In addition,
from a national library perspective, to know the geographical distribution
of place names in its own territory is of greater interest than to
know the mention of foreign place names. Finally, smaller populated
places, such as neighbourhoods, have been left out after some preliminary
tests indicated the presence of noise.

Priority has been given to a method offering the best balance between
recall and precision for these places. This method is based on an
initial list of some 3000 Belgian municipalities and municipal districts
which we have semi-automatically enriched by matching these names
with several KBs. For every location, our authority file contains additional information
such as its Wikipedia page (in French, Dutch or English), its Wikidata
ID, geographical coordinates, etc. Most importantly, this file is
linked to a second table containing all the aliases and alternative
spellings mentioned in the GeoNames dumps\footnote{\url{http://download.geonames.org/export/dump/}},
an extensive open geographic database containing more than 11 million
places around the world. For example, the file contains no less than
119 various spellings and language versions of the Flemish city of
Antwerp.

The extraction of place names consists of retrieving in the queries
any presence of one of these spellings. Once the query is subdivided
in tokens, the central task is to match each token (or as many tokens
as possible) with one of the location names contained in an authority
file. This step allows us to match isolated tokens (such as \enquote{brussels}
with \enquote{Bruxelles}) as well as more complex queries (such
as \enquote{mont sainte aldegonde} with \enquote{Mont-Sainte-Aldegonde},
which was written without hyphens in the original query and is therefore
separated in different tokens). Finally, we used a list of common
names in French and Dutch, as well as a list of first names (section
\ref{subsec:Personal-names}), to develop some minimal linguistic
rules to minimise for example possible confusions between the first
name \enquote{Hervé} and the Belgian municipality
\enquote{Herve}).

In order to avoid an excess of false recognition and to limit the
processing time, we have decided not to use a fuzzy matching algorithm.
Geographic references therefore have to be written in a strictly identical
manner in order to be recognised. Thus, \enquote{Anwterp}, a misspelled
mention of the Flemish city, will not be matched with the correct
spelling \enquote{Antwerp}. 

\emph{\subsection{\label{subsec:Personal-names}Personal Names}}

The workflow to automatically extract personal names is similar, except
that in this case we wanted to promote the recall rather than the
precision. Indeed, these candidate names are intended to be verified
by using KBs, which will ultimately sort out false
positives. Postulating that a first name or a last name alone is not
sufficient to identify a person\footnote{Except for the name of well-known figures such as Hergé or La Callas,
which we choose to omit in this prototype.}, we decided to start by retrieving a maximum of possible full names
(first name and last name).

The first step to process queries and identify those containing a
full name of person consists of isolating every query composed of
more than one token. All potential full names will be in the resulting
subset (\enquote{candidate names}). The method consists in using
a \enquote{trigger} word, in this case the presence of a first
name. We matched first names mentioned in the queries with an authority
file containing nearly 25 000 given names extracted from Wikidata
\footnote{\url{https://www.wikidata.org/wiki/Wikidata:WikiProject_Names/lists/given_names_by_soundex}}.
Then, the aim is to deduce, with the aid of a small set of linguistic
rules, which part of the query probably constitutes the last name.
Attention has been paid to include names involving a particle (e.g.
van, von, de, van den, van der, ...) and other ones composed by more
than two tokens. 

The second step aims to link these candidate names to KBS which are likely to identify them. This operation, called Named Entity Linking \citep{rao2013entity}, goes beyond the NERC and encompasses it --- all the more so that both can be mutually reinforcing \citep{Luo2015, cornolti2016piggyback}. Two KBs have been used in this project: a general one, Wikidata, and a more specialised, VIAF: 

\begin{itemize}
\item {Wikidata, launched in 2012, is a data directory intended to feed sister Wikimedia projects, especially the information boxes of almost 300 linguistic editions of Wikipedia. Wikidata offers many advantages over other similar KBs such as DBpedia or Yago, as it includes more frequent updates and a greater number of entities \citep{Geiss2017}. In August 2017, this KB comprised almost 34 million entities \citep{Statisti56}, of which about 10\% of human beings --- the major category. 
Unlike other KBs such as DBpedia, Wikidata is language-independent: each concept or property is being referenced by a unique URI. In addition, its data can be edited by both humans and machines and are not limited to those extracted from Wikipedia. Thus, since 2016, Wikidata is gradually integrating the extensive Freebase knowledge
base, previously owned by Google. 

\item {VIAF \emph{(Virtual International Authority File)}, hosted since 2012 by the OCLC, seeks to link authority files owned by cultural heritage institutions (originally national libraries, but nowadays also museums or Wikipedia) and to make them available online. Beside its founding members, the Library of Congress and the Deutsche Nationalbibliothek, the consortium includes today 37 partner institutions from 29 countries. In 2016, its statistics reported 33 million cluster records, of which about 7.5 million were names of persons \citep{VIAFRefl77}. It should be mentioned that, in comparison with Wikidata, VIAF contains nearly twice as many personal names}}
\end{itemize}

\noindent These two free KBs have been used for the reconciliation. Unlike the extraction of Belgian place names, based on a closed file which has been enriched a posteriori, the recognition of personal names relies on the web services of Wikidata and VIAF and their free APIs. The discussion on the architecture and constraints of APIs and their conformity to the REST principles, falls outside the scope of the paper but \citet{verborgh2015fallacy} offer an in-depth overview of these issues.   
These APIs allowed us to create a subset among the candidate names with all the person names possessing a VIAF or a Wikidata entry. Remaining entities include both those which can be considered as unknown but existing people (e.g. Adeline Pollet) and those which are extraction
errors (e.g.\,\enquote{Pole Nord}, which means \enquote{North Pole} in French).

\subsection{Evaluation}

Before applying our method to all queries (see \ref{sec:Results}), we evaluated it on our annotated sample ({GSC} as a primary evaluation, see \ref{sec:section5}). In order to be aligned with the rules of our extractor, we have adapted this
GSC by adding two new subsets. For place names (LOC), we created a subset containing only Belgian municipalities and municipal districts.
For personal names (PER, PER LOC and PER AMBIG), we created a subset containing exclusively (supposed) full names, i.e. entities composed by a combination of first name and last name.

\begin{figure}
\begin{raggedleft}
\includegraphics[scale=0.50]{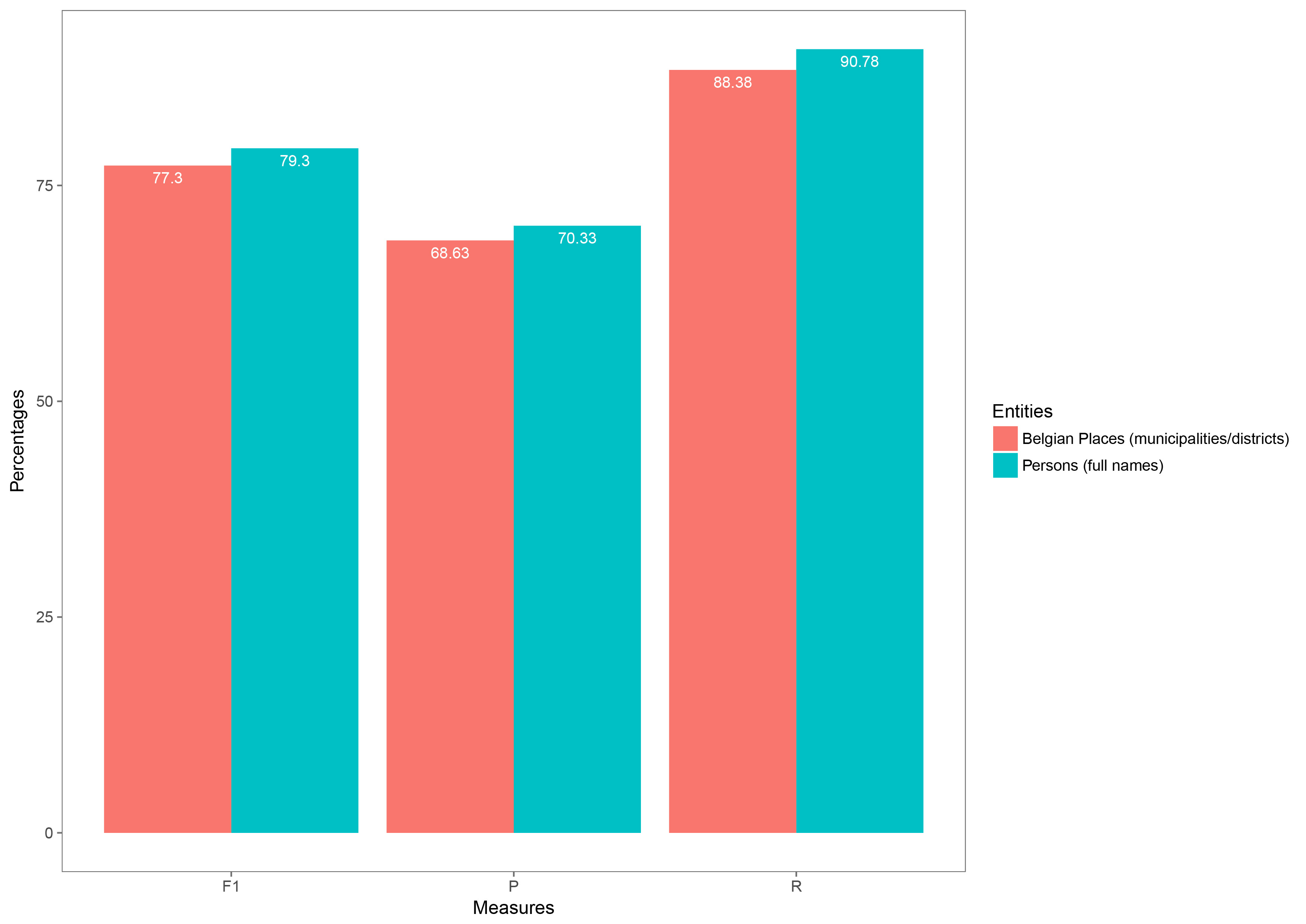}
\par\end{raggedleft}
\caption{\label{fig:Evaluation-results-in}Evaluation results in terms of Recall,
Precision and $ F_{1} $-score}

\end{figure}

For place names, our script has correctly extracted 175 municipalities
and municipal districts out of 198 reference entities. The main omissions
are due to abbreviations (\enquote{wez} for \enquote{Wez-Velvain})
or barely recognisable spellings (\enquote{overrvssche} for the
municipality of \enquote{Overijse}).
For personal names, our script has correctly extracted 128 full names
out of 141 reference entities. Beside these 128 entities which are
strong annotation matches, the script has extracted 47 false positives
and 7 incomplete matches (for example \enquote{loo prosper} instead
of \enquote{Van Loo Prosper}). 
These results are summarised in Figure \ref{fig:Evaluation-results-in} using the standard measures of Precision\footnote{The percentage of correctly identified Named Entities in all Named Entities extracted.}, Recall\footnote{The percentage of Named Entities found compared to all existing Named Entities.} and $ F_{1} $ score\footnote{The weighted harmonic mean of precision and recall, with in this case
equal weights for both.}.

\subsection{\label{sec:Results}Results}
Applied to the entire corpus, our tool returned 16 670 mentions of Belgian municipalities and municipal districts, and 13 463 mentions of full personal names. 

\begin{figure}
\begin{centering}
\includegraphics[scale=0.30]{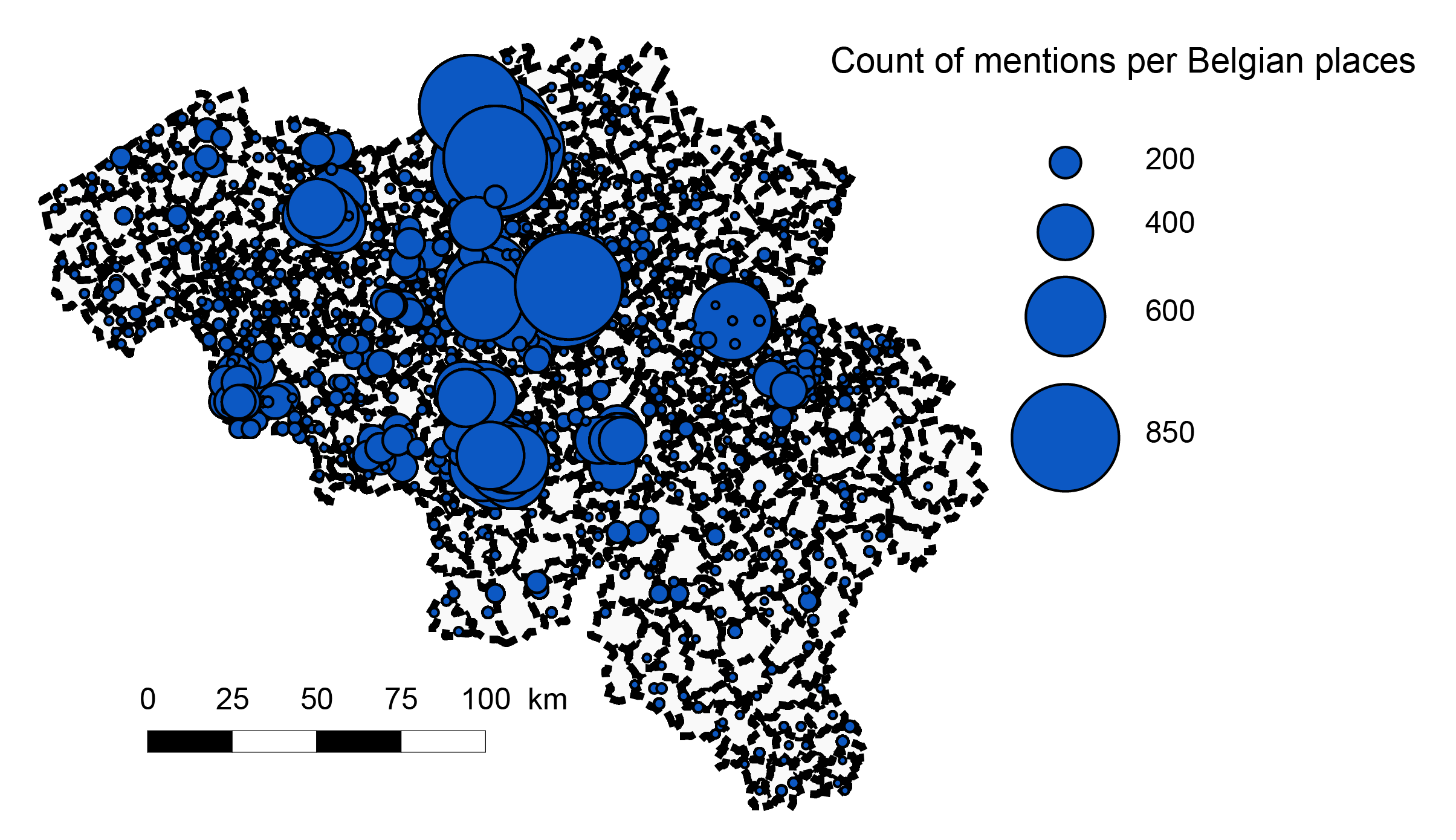}\caption{\label{fig:Mapping}Mapping of the number of mentions per municipality.}
\par\end{centering}
\end{figure}

The extraction of place names can be of particular
interest to the collection holder. As shown in Figure \ref{fig:Mapping}, this operation allows to visualise the possible imbalances in the geographical distribution of the area's of interest. If necessary, the script also allows to weigh the number of queries associated with a place by the population of the municipality to which it belongs.
We will limit ourselves to pointing out that the distribution between municipalities and districts is more unbalanced than in the sample, since the municipalities extracted represent 59\% of the total corpus.

\begin{figure}
\centering{}\includegraphics[scale=0.4]{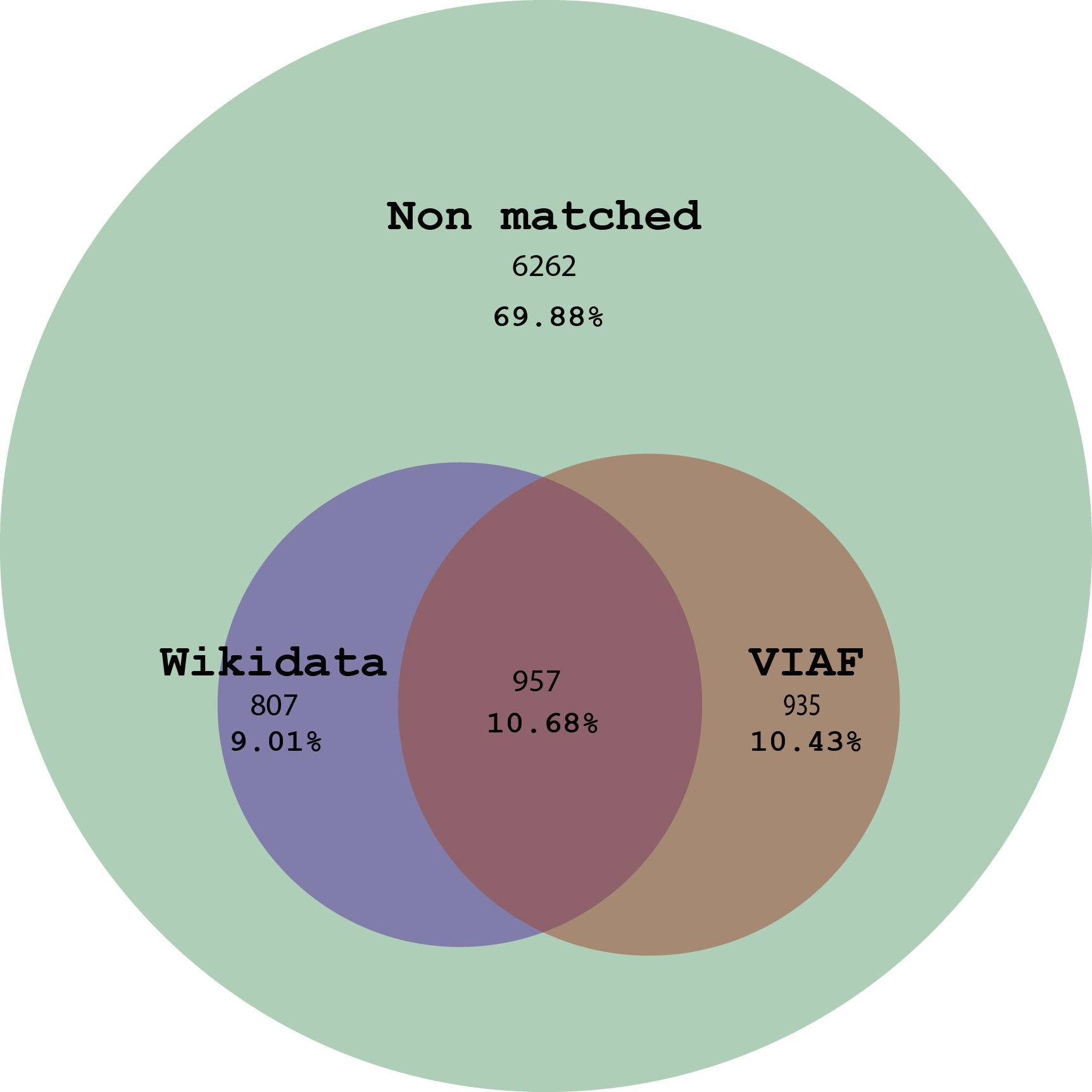}\caption{\label{fig:Repartition}Breakdown by number and percentage of total
candidate names (diagram to scale)}
\end{figure}

In the case of personal names, the 13 463 mentions correspond to 8
961 different spellings, which will be considered as so many different
persons. Among all these candidate names, 2 699 (30.1\%) could be
matched by at least one KB. Figure \ref{fig:Repartition}
shows the distribution between VIAF and Wikidata. About one third of personal names have been retrieved uniquely in VIAF, one other third only in Wikidata, the rest appearing in both. These results underline the complementarity of using both KBs. 

\section{Discussion}
This case study has shown how the exploration of user needs through
user queries can be enhanced and semi-automated by making use of NLP
and KBs published as Linked Data such as Geonames, Wikidata or VIAF. However, the
satisfying F-score of our extractor aside, one must recognise some
limits.

On the one hand, there are limits inherent to the queries themselves.
The major difficulty is related to lexical ambiguity, which is accentuated
by the lack of context of user queries. The homonymy concerns both
place names and personal names. For example, several municipalities
possess the same name. Thus, a query containing the place name \enquote{Saint-Nicolas}
could be associated with different Belgian municipalities called \enquote{Saint-Nicolas}
or even \enquote{Sint-Niklaas}. The problems of homonymy stretch
even further: we cannot know if the user was looking for a village
called Saint-Nicolas or festivities and traditions of December 6,
related to Saint Nicolas. The same problem arises with personal names.
For example, first names can also be common names, such as \enquote{Fleur},
which can be a proper name or the French translation of flower. Dealing
with this lexical homonymy requires vigilancy to identify problematic
cases, to conduct cost-benefit analyses and to set lexical rules to
limit noise while maintaining recall. Finally, although contextual
elements such as other queries performed during the same visit could
be used to disambiguate a query, there is no way of knowing which
person was really designated, except by diving into the end user's
mind.

On the other hand, some limits are related to the authority data used
to retrieve NE. First, processes and results rely on the
quantity of data made available online. Figure 3 demonstrates how the percentage of retrieved personal names crucially depends on the size and the scope of KBs used. Let us illustrate this with an example: \enquote{Lodewijk Vander Schopen}, a Belgian writer of the 19th century. Fortunately, the Dutch National Library shares its data with VIAF, making it possible to identify this individual.\footnote{\url{http://VIAF.org/VIAF/286861725/}} 
If this were not the case, Lodewijk Vander Schopen would not have been retrieved, given that the Dutch National Library is the only institution mentioning him in VIAF and that no Wikidata entry refers to this writer. This example, as well as the overlap between Wikidata and VIAF being only about one-third, emphasises how these knowledge bases complement each other. It also highlights the strategic importance of the type of Linked Data which will be used for the data reconciliation. 

Finally, some limitations are directly related to the process of matching queries and data in KBs. Each API has its own rules for assigning a score to candidate entities, and these rules are sometimes poorly documented. In the case of Wikidata, this is a purely approximate string matching with the different spellings and \enquote{also known as} listed in the database. Thus the \enquote{Curer Bell} query could be reconciled with the British poet and novelist Charlotte Brontë \footnote{\url{https://www.wikidata.org/wiki/Q127332}}, of which it is one of the pseudonyms. The two names will be associated with a score of more than 90\% probability, the threshold which we retained for the matching. The VIAF API, however, takes into account more parameters for the matching process. Its proposals are based not only on string matching with the names recorded by the various partner libraries, but also with other metadata contained in the record of each personality. Thus,
for the request \enquote{Victor Guillemin}, VIAF
proposes as match \enquote{Hugo, Victor, 1802-1885},
which appears to be irrelevant. The explanation is simple: the French writer's
page \footnote{\url{https://viaf.org/viaf/9847974/\#Hugo,_Víctor,_1802-1885}}
mentions among his \enquote{related names} a certain \enquote{Guillemin, Henri}. The strings
\enquote{Victor} and \enquote{Guillemin}
are therefore associated. Another example is the query \enquote{Guillaume
Archiduc}, for which Wikidata does not return any results.
VIAF proposes matches with two different individuals (\enquote{Habsburg-Lothringen, Wilhelm, 1895-1949} and \enquote{Leopold Wilhelm, Archduke of Austria, 1614-1662}): both are archdukes and both are named, among others, Guillaume. This time, the API provides relevant results. As this example indicates, VIAF retrieves results of equivalent names across languages, even though the names are strongly different. However, both retrieved persons are from different centuries, which indicates the lack of guarantee of relevancy.

\section{Conclusions and Future Work}
As demonstrated in the discussion of the results, our method of matching user queries
with place and person names provides salient results.
Given that it is freely available for other cultural heritage institutions,
the tool is perfectly suited to assist other Belgian institutions to perform a similar analysis,
and sufficiently generalisable to be customised by libraries, archives and museums outside Belgium.
For the identification of place names, the Geonames values specific to the country should be used.
Regarding the identification of person names, an institution should identify its own local authority list of first names and configure the list with tokens typically included in family names, such as \enquote{van} or \enquote{von} for Dutch or German. 

Even if the technology is easily accessible, this paper underlined the importance of a lengthy iterative process of precise adjustments
which influence the results to a large extent. As demonstrated with the matching process
based on VIAF for example, our results would be very different if we had allowed a matching based on a greater spelling
difference between the query and the candidates it proposed.
Far from the simplistic image that it merely takes a few clicks to find
all the names of people or places in the queries, the paper demonstrated with concrete examples
the complexity of the process. The process involves a set of successive choices, each one requiring a cost-benefit
analysis concerning the delicate balance to be found between optimising either precision or
recall. 

Conceptually speaking, this paper also sheds a more nuanced light on how we can leverage KBs published as Linked
Data for documentation practices. Within this rapidly evolving landscape, demonstrated recently by the phasing out of Freebase and the rapid rise of Wikidata for example, it can be hard to understand which environments can be considered as a valid and sustainable source of authority files.
As demonstrated in an analysis of the triples on the topic of Henry IV, van Hooland and Verborgh (2014) demonstrated
how various KBs can offer very different metadata on the French king, who notoriously swapped religions for pragmatical political reasons.
The Linked Data community often underlines how competing KBs are all interconnected through predicates such as \emph{owl:sameAs}, but when operationalising these links, divergent and sometimes conflicting metadata often come to the surface.  

Also, the quantitative analysis clearly revealed how a general KB (Wikidata) can be complemented by  making use of VIAF, specifically geared at identifying and disambiguating names of authors.  
These APIs allowed us to create a subset within the candidate names with all the person names possessing a VIAF or a Wikidata entry. Remaining entities include both those which can be considered as unknown but existing people (e.g. Adeline Pollet) and those which are extraction
errors (e.g.\,\enquote{Pole Nord}, which means \enquote{North Pole} in French).
Based on these data, one could potentially also infer how many queries refer to unknown people, and which are of interest to genealogists, in regards to the amount of queries on famous historical figures, of interest to researchers. These are typical research questions
for which we can operationalise Linked Data, taking into account the caveats highlighted across our paper.

In terms of future work, we wish to extend the current approach based on string matching. The next step would be to rank these candidates in order of probability. This disambiguation process requires a minimal context around the queries. One of the possibilities would be to use the other queries performed by a user during the same visit and to compare the set with a large textual database. Attempts made with Wikipedia on our annotated sample proved unsuccessful. Indeed, too many names of people searched in BelgicaPress are totally unknown to Wikipedia. On the other hand, other tests using the commercial Google Books API seem promising. Other options include using the pretrained Word2Vec algorithm to assign similar words to the queries, and by doing so, augmenting the success rate of the matching process.  

\section{Acknowledgements}

The authors would like to extend their gratitude to the Royal Library of Belgium. They are particularly grateful for the assistance given by Erwin Van Wesemael. The support of the promotors of the ADOCHS project has also been invaluable to the success of the research and the conception of this article. The authors would therefore like to thank Ann Dooms, Florence Gillet and Frederic Lemmers. The research underlying the results presented in this article was funded by the Belgian Science Policy Office in the context of contract number BR/154/A6/ADOCHS.

\theendnotes

\bibliographystyle{agsm}
\bibliography{references2}

\end{document}